\PassOptionsToPackage{table,xcdraw}{xcolor}
\documentclass[11pt,letterpaper]{article}
\usepackage[margin=1in]{geometry}
\usepackage{amsmath,epsfig}
\usepackage{stfloats}
\usepackage{breqn}
\usepackage{booktabs}
\usepackage{multirow}
\usepackage{multicol}
\usepackage{pdflscape}
\usepackage{array}
\usepackage{xcolor}
\usepackage{colortbl}
\usepackage{caption}
\usepackage{makecell}
\usepackage{xcolor}
\usepackage{float}
\usepackage{hyperref}
\usepackage{url}
\usepackage{subcaption}
\usepackage{amsmath,amssymb,amsfonts}
\usepackage{graphicx}
\usepackage{booktabs}
\usepackage{caption}
\usepackage[style=ieee]{biblatex}

\definecolor{lightgray}{gray}{0.9}

\def\X{{\mathbf X}}

\newcolumntype{M}{>{\centering\arraybackslash}m{2.1cm}}  % method column
\newcolumntype{R}{>{\centering\arraybackslash}m{1.1cm}}  % (r,c) column
\newcolumntype{D}{>{\centering\arraybackslash}m{1.4cm}}  % dataset column

\title{HyperBench: Standardizing and Scaling Synthetic Evaluation\\ for Hyperspectral Super-Resolution}

\author{
  Ritik Shah\\
  University of Massachusetts\\
  Amherst, MA 01003\\
  \texttt{rgshah@umass.edu}
  \and
  Marco F. Duarte\\
  University of Massachusetts\\
  Amherst, MA 01003\\
  \texttt{mduarte@umass.edu}
}

\date{}

\begin{document}

\maketitle

\begin{abstract}
Hyperspectral super-resolution (HSR) reconstructs a high-spatial-resolution hyperspectral image by fusing a low-resolution hyperspectral image (LR-HSI) with a high-resolution multispectral image (HR-MSI). In the absence of real-world paired data, HSR methods are evaluated almost exclusively on synthetic experiments derived from hyperspectral datasets through Wald's protocol. Despite the protocol's widespread adoption, its practical implementation varies markedly across research works, typically relying on a single (usually Gaussian) or very few point spread functions (PSFs), one or two spectral response functions (SRFs), and a couple of spatial downsampling factors. As a result, reported performance figures are difficult to compare across the literature, in addition to being often difficult to reproduce; furthermore, they may not generalize across realistic sensing conditions.

We introduce \textbf{HyperBench}, a unified and extensible framework that standardizes synthetic experimentation for HSR. HyperBench supports diverse degradation configurations spanning ten PSFs, four SRFs derived from operational multispectral sensors, configurable spatial downsampling factors, and matched additive white Gaussian noise; its goal is to automate large-scale evaluation and structured logging. By decoupling model development from experimental design, the framework enables reproducible, apples-to-apples cross-method comparison with minimal friction. We use HyperBench to evaluate six recently proposed HSR methods across a 70-configuration sweep on four widely used hyperspectral scenes and observe that the inter-method PSNR spread widens from approximately $5$\,dB on the easiest PSF to over $13$\,dB on the hardest---a fragility that is structurally invisible to the prevailing single-configuration evaluation protocol. HyperBench code is available at \url{https://github.com/ritikgshah/HyperBench} and can be downloaded using pip install hyperbench.
\end{abstract}

\section{Introduction}

Hyperspectral imaging enables detailed characterization of scenes by capturing dense spectral information across hundreds of contiguous wavelength bands. This capability is critical in applications such as remote sensing, environmental monitoring, and material identification. However, hyperspectral sensors are fundamentally limited in spatial resolution due to hardware and physical constraints. In contrast, multispectral sensors provide higher spatial resolution but capture significantly fewer spectral bands. Hyperspectral super-resolution (HSR) addresses this trade-off by fusing a low-resolution hyperspectral image (LR-HSI) with a high-resolution multispectral image (HR-MSI) to reconstruct a high-resolution hyperspectral image (HR-HSI).

Formally, the HSR problem seeks to recover an image of shape $(H, W, C)$ from inputs of shapes $(h, w, C)$ and $(H, W, c)$, where $H/h$ and $W/w$ denote spatial scaling factors and $c \ll C$. In the absence of real-world paired datasets, evaluation of HSR methods relies almost exclusively on synthetic experiments derived from hyperspectral datasets using Wald’s protocol \cite{wald}. Under this protocol, a hyperspectral image is treated as ground truth and degraded to simulate sensor observations, producing paired LR-HSI and HR-MSI inputs.

While Wald’s protocol provides a practical and widely accepted evaluation paradigm, its implementation in the literature has historically been narrow and inconsistent. Most prior works adopt simplified degradation models, typically relying on Gaussian point spread functions (PSFs) which only capture a small subset of blurring that may occur in practice in hyperspectral sensors, a small number of downsampling factors, and limited spectral response functions (SRFs) \cite{HySure,CNMF,MIAE,SDP}. These choices significantly restrict the diversity of evaluation conditions. As a result, many methods are implicitly optimized for a narrow set of assumptions, and their reported performance may not generalize beyond these settings.

In the course of evaluating new methods \cite{SpectraLift,SpectraMorph} against prior state of the art, we observed that reported HSR performance depends heavily on which PSFs and SRFs are chosen. Methods that claimed state-of-the-art results while reporting on only a narrow set of degradation operators often did not retain those gains when re-evaluated under alternative PSFs or SRFs. Several published HSR models report metrics only for a Gaussian PSF, some omit spatial blurring altogether despite its physical presence in hyperspectral sensors, and others omit noise. Table~\ref{tab:eval-survey} summarizes the synthetic evaluation conditions reported by a representative selection of HSR methods, making the narrowness of the prevailing protocol concrete.

\begin{table}[t!]
\centering
\small
\caption{Reported synthetic evaluation conditions in representative HSR works. Each row reflects the configuration documented in the cited paper. Most works (top) fix a single Gaussian PSF, a single SRF, and a single downsampling factor; only recent investigations (bottom) sweep over multiple PSFs, SRFs, and scales.}
\label{tab:eval-survey}
\begin{tabular}{|l|l|c|c|l|}
\hline
Method & PSF used & \# SRFs used & Down. factor & Noise modeled \\
\hline
HySure \cite{HySure} & 3 distinct kernels & 2 & 4 & Gaussian\\
\hline
CNMF \cite{CNMF} & Gaussian & 1 & 6 & Gaussian \\
\hline
MIAE \cite{MIAE} & Gaussian & 1 & 8 & Gaussian \\
\hline
C2FF \cite{C2FF} & Gaussian & 1 & 12 & Gaussian \\
\hline
SDP \cite{SDP} & Gaussian & 1 & 8 & Gaussian \\
\hline
SSSR \cite{SSSR} & Gaussian & 2 & 8 & None mentioned \\
\hline\hline
SpectraLift \cite{SpectraLift} & 10 distinct kernels & 5 & 4, 8, 16, 32 & Gaussian \\
\hline
SpectraMorph \cite{SpectraMorph} & 10 distinct kernels & 5 & 4, 8, 16, 32 & Gaussian \\
\hline
\end{tabular}
\end{table}

The expanded experimentation strategies introduced in the works above remain tightly coupled to their specific implementations. Reproducing or extending these experiments to additional methods, configurations, or scenes requires substantial manual effort, with the result that evaluation practice in HSR remains fragmented, difficult to reproduce, and challenging to scale. In contrast, much of the research in data-centric disciplines has observed the emergence of {\em frictionless reproducibility}, which describes the fast progress of research as being driven by three pillars: making data freely shareable, providing easily re-executable code, and emphasizing competitive benchmarking as a mode of evaluation. Together, these practices allow research to be replicated at the press of a button, dramatically lowering the barriers to building on others' work and accelerating the pace of discovery~\cite{donoho2024datascience}.

In this work, we introduce \textbf{HyperBench}, a framework that provides frictionless reproducibility for synthetic experimentation in HSR research. HyperBench standardizes the generation of synthetic observations under Wald's protocol, provides a configurable degradation library encompassing ten PSFs, four SRFs derived from operational multispectral sensors, arbitrary spatial downsampling factors, and matched additive noise, and automates the execution and evaluation of reconstruction methods across user-specified parameter grids. By decoupling model development from experimental design, the framework makes broad, reproducible, and apples-to-apples cross-method comparison computationally tractable rather than manually prohibitive.

To demonstrate the framework's utility, we use HyperBench to evaluate six recent HSR methods across $70$ degradation configurations on four widely used hyperspectral scenes (Section~\ref{sec:study}). The resulting cross-method robustness profile shows that the inter-method PSNR spread varies by more than threefold across the PSF axis alone---widening from approximately $5$\,dB on the easiest PSF to over $13$\,dB on the hardest---and that several methods reported as state-of-the-art under the conventional single-Gaussian protocol lose $8$--$12$\,dB under alternative spatial degradations. These findings make a concrete case for adopting standardized, multi-configuration evaluation as a baseline practice in the field, and are themselves possible because HyperBench reduces such studies from a per-paper engineering undertaking to a single configuration file.

\section{Synthetic Evaluation under Wald’s Protocol}

Let $X \in \mathbb{R}^{H \times W \times C}$ denote a hyperspectral image. Under Wald’s protocol, this image serves as a reference from which degraded observations are generated.

\subsection{Ground Truth Construction}

To ensure numerical stability and consistency across datasets, the input hyperspectral image is normalized using percentile-based clipping. The ground truth $X_{GT}$ is defined as:

\begin{equation}
X_{GT} = \frac{\mathrm{clip}(X, p_l, p_u) - p_l}{p_u - p_l}
\end{equation}

where $p_l$ and $p_u$ denote lower and upper (1 and 99) percentiles. This normalization reduces the influence of outliers and ensures a consistent dynamic range.

\subsection{Generation of LR-HSI}

The low-resolution hyperspectral image $X_{LR} \in \mathbb{R}^{h \times w \times C}$ is obtained through spatial degradation:

\begin{equation}
X_{LR} = D(B(X_{GT})) + N_H
\end{equation}

where $B(\cdot)$ represents convolution with a point spread function (PSF), $D(\cdot)$ denotes spatial downsampling such that $H/h = W/w$, and $N_H$ is additive white Gaussian noise. This process simulates the spatial limitations of hyperspectral sensors.

\subsection{Generation of HR-MSI}

The high-resolution multispectral image $Y_{HR} \in \mathbb{R}^{H \times W \times c}$ is generated via spectral degradation:

\begin{equation}
Y_{HR} = S(X_{GT}) + N_M
\end{equation}

where $S(\cdot)$ is a spectral response function mapping $C$ spectral bands to $c \ll C$ bands, and $N_M$ is additive white Gaussian noise.

\section{HyperBench Framework}

HyperBench resolves a central bottleneck in HSR research by standardizing synthetic experimentation. Rather than treating data generation, experiment design, and evaluation as ancillary scaffolding around each new model, the framework consolidates them into a single, structured pipeline. Given a hyperspectral dataset, HyperBench constructs a normalized ground truth, generates LR-HSI and HR-MSI observations under configurable degradation operators, executes user-defined super-resolution methods, and evaluates their outputs against the ground truth. Because the entire pipeline is fixed and shared, methods are inserted as modular components: users provide a single model interface, and HyperBench handles all surrounding mechanics. Reported performance differences across methods are therefore traceable to model differences rather than to incidental implementation choices.

The framework supports both large-scale and targeted experimentation. In the large-scale mode, users specify sets of PSFs, SRFs, downsampling factors, and SNR levels; HyperBench then evaluates every method across the Cartesian product of these sets, effectively conducting a structured grid search over the degradation space. This mode makes systematic and comprehensive robustness studies frictionless. In the targeted mode, users specify individual configurations and execute experiments through a single function call, supporting controlled ablations, debugging, and reproduction of prior work.

All experiments are evaluated and logged automatically. For each run, HyperBench computes a configurable set of reconstruction metrics and records both the metric values and the full configuration context (PSF parameters, downsampling factor, SNR levels, and random seeds) in a structured tabular format. This makes large experimental suites trivially aggregatable and enables direct comparison across methods, configurations, and datasets without per-paper bookkeeping.

\section{Degradation Models}

A central design goal of HyperBench is to move beyond narrow sets of degradation assumptions that have common in synthetic evaluation in HSR. In practice, real hyperspectral sensors exhibit a diverse and complex range of spatial and spectral distortions that are poorly approximated by a single Gaussian blur and a fixed spectral response. HyperBench standardizes this diversity by providing a modular and extensible library of degradation models that can be composed into reproducible experimental configurations.

\subsection{Point Spread Functions}
\label{subsec:psf}

The spatial degradation applied to generate the LR-HSI is governed by a point spread function (PSF), which models the optical blur introduced by a hyperspectral sensor's optics and detector array. The choice of PSF critically influences the difficulty and character of the super-resolution problem, yet most prior works default to an isotropic Gaussian kernel, leaving model robustness under alternative blur conditions unexamined. HyperBench incorporates ten distinct PSFs spanning a range of isotropic and anisotropic
blur profiles. The supported PSFs are:

\begin{itemize}
    \item \textbf{Gaussian}~\cite{Goodman2005}: A smooth, isotropic kernel widely used as a baseline in the literature. While convenient, it does not capture the structured or asymmetric blurring commonly observed in real sensors.

    \item \textbf{Kolmogorov}~\cite{Fried1966}: Derived from atmospheric turbulence theory, this PSF models the blur induced by atmospheric propagation effects relevant in airborne and satellite sensing.

    \item \textbf{Airy}~\cite{Airy1835}: Arising from diffraction-limited optics, the Airy PSF introduces a characteristic ring structure that is physically grounded in the wave optics of circular apertures.

    \item \textbf{Moffat}~\cite{Moffat1969b,buonanno1988ccd}: Originally developed for astronomical imaging, the Moffat PSF provides a heavier-tailed alternative to the Gaussian, better capturing extended scattering profiles.

    \item \textbf{Sinc}~\cite{Shannon1949}: The ideal low-pass filter kernel, associated with perfect bandlimited interpolation. Its oscillatory sidelobes present a distinct challenge for reconstruction methods.

    \item \textbf{Lorentzian Squared}~\cite{Moffat1969}: A heavy-tailed, peaked kernel that introduces pronounced central sharpness combined with broad energy spread, testing reconstruction fidelity across spatial frequencies.

    \item \textbf{Hermite}~\cite{Martens1990}: A kernel derived from Hermite polynomial modulation of a Gaussian envelope, introducing structured oscillations that simulate complex optical point responses.

    \item \textbf{Parabolic}~\cite{Epanechnikov1969}: A compactly supported kernel with smooth, symmetric spatial falloff, providing an intermediate regime between diffraction-limited and geometric blur models.

    \item \textbf{Gabor}~\cite{Gabor1946}: An anisotropic kernel combining a sinusoidal carrier with a Gaussian envelope, capturing directional blur arising from sensor motion or optical misalignment.

    \item \textbf{Delta Function}: The identity kernel that applies no spatial blur, serving as a controlled baseline for isolating the effect of downsampling and noise in isolation from spatial blurring.
\end{itemize}

By evaluating models across this range of PSFs, HyperBench exposes behavioral differences that remain invisible under Gaussian-only evaluation. In particular, models that are implicitly overfit to smooth, isotropic blur may fail to generalize to anisotropic or oscillatory kernels. Making this failure mode explicit is a prerequisite for developing methods that are genuinely robust to realistic sensing conditions.

\begin{figure*}[!ht]
    \centering
    \includegraphics[width=\textwidth]{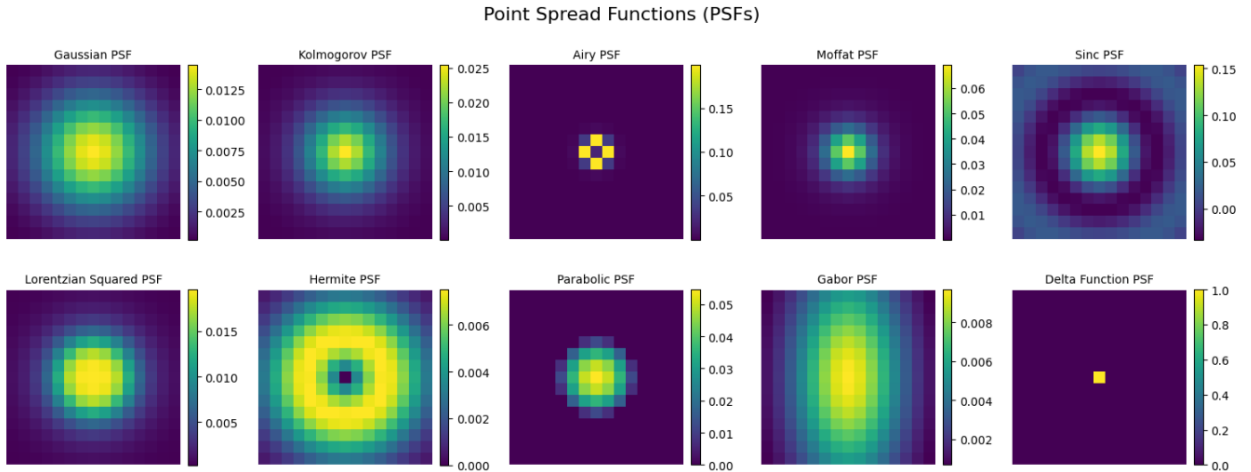} 
    \caption{Point Spread Functions supported with HyperBench}
    \label{fig:psf}
\end{figure*}

\subsection{Spectral Response Functions}
\label{subsec:srf}

The spectral degradation applied to generate the HR-MSI is governed by a spectral response function (SRF), which maps the full $C$-band hyperspectral image to a lower-dimensional multispectral observation with $c \ll C$ bands. The SRF encodes the spectral sensitivity profile of the companion multispectral sensor and has a direct bearing on how much spectral information is preserved in the HR-MSI input.

Despite this significance, the majority of published HSR methods are evaluated against a single SRF derived from one specific sensor configuration. This practice severely limits the generalizability of reported conclusions, as a method that leverages the specific spectral correlations present in one SRF may offer no advantage under a different spectral decomposition.

HyperBench addresses this by providing SRFs derived from four representative sensor platforms covering 3, 4, 8, and 16 spectral bands, spanning the visible through shortwave infrared (SWIR) spectrum. The 3- and 4-band SRFs are based on IKONOS-style configurations \cite{DigitalGlobeSRF}, capturing the RGB and RGB+NIR sensitivities typical of commercial multispectral cameras. The 8-band SRF is based on WorldView-2 \cite{DigitalGlobeSRF}, representative of multispectral systems used in aerial and satellite remote sensing. The 16-band SRF is based on WorldView-3 \cite{DigitalGlobeSRF}, which extends coverage into the SWIR. Together, these four configurations represent a meaningful cross-section of the operational sensor landscape, enabling evaluation under varying degrees of spectral resolution and wavelength coverage.

\begin{figure*}[!ht]
    \centering
    \includegraphics[width=\textwidth]{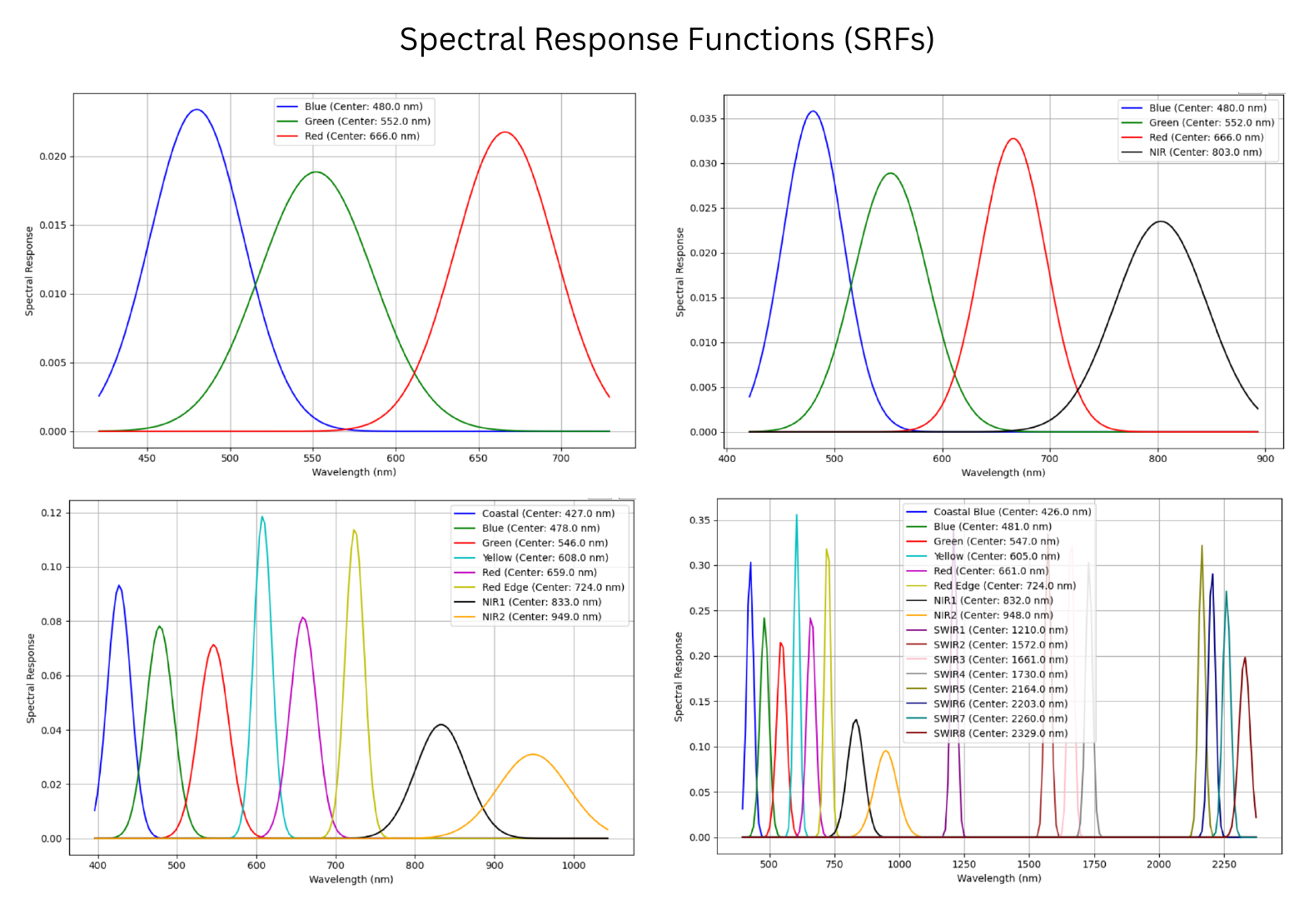} 
    \caption{Spectral Response Functions supported with HyperBench}
    \label{fig:srf}
\end{figure*}

\subsection{Downsampling Factors and Noise}

In addition to PSF and SRF variability, HyperBench supports configurable spatial downsampling factors governing the ratio $H/h = W/w$ between the HR-HSI and LR-HSI resolutions. The choice of downsampling factor directly determines the difficulty of the super-resolution task and influences the relative contribution of spatial versus spectral information to the final reconstruction. HyperBench allows users to specify arbitrary integer downsampling factors using OpenCV area interpolation, enabling systematic evaluation across resolution regimes.

Additive white Gaussian noise is applied independently to both the LR-HSI and HR-MSI observations at configurable signal-to-noise ratios. This model is well-suited to the high-photon-count and electronic-noise regimes that dominate most operational hyperspectral acquisitions, where photon-shot noise is well approximated by a Gaussian distribution. In photon-starved regimes, however, the noise is more accurately described by a Poisson process; HyperBench currently models noise as Gaussian only, and Poisson noise is planned as a future addition through the same modular interface. Support for noise-free configurations is also provided, allowing controlled isolation of the effects of spatial and spectral degradation from noise.

Collectively, the full degradation pipeline of HyperBench---spanning PSF type, downsampling factor, SRF configuration, and noise level---defines a structured evaluation space that is orders of magnitude larger than what any individual prior work has explored. The framework makes this space tractable through automated grid evaluation, transforming comprehensive benchmarking from an exceptional undertaking into a routine part of the research workflow.

\section{Evaluation Metrics}

To comprehensively assess reconstruction quality, HyperBench computes six widely adopted metrics that capture complementary aspects of reconstruction quality in hyperspectral super-resolution:

\begin{itemize}
    \item {\bf Root Mean Squared Error (RMSE):} Measures the average pixel-wise difference between the reconstructed hyperspectral image \(\hat{\X}\) and the ground truth \(\X\). Lower RMSE indicates higher reconstruction fidelity.
    \[
    \mathrm{RMSE}(\X,\hat{\X}) = \sqrt{\frac{1}{HWC}\sum_{i=1}^{H}\sum_{j=1}^{W}\sum_{k=1}^{C} \bigl(\hat{\X}_{ijk} - \X_{ijk}\bigr)^2}.
    \]

    \item {\bf Peak Signal-to-Noise Ratio (PSNR):} Quantifies the ratio between the maximum possible pixel value and the power of the reconstruction error, expressed in decibels (dB). Higher PSNR indicates better perceptual quality.
    \[
    \mathrm{PSNR}(\X,\hat{\X}) = 10 \log_{10}\left(\frac{\mathrm{MAX}^2}{\mathrm{RMSE}(\X,\hat{\X})^2}\right),
    \]
    where \(\mathrm{MAX}\) is the maximum possible pixel value (e.g., \(2^{16}-1\) for 16-bit images).

    \item {\bf Structural Similarity Index Measure (SSIM)}~\cite{SSIM}: Assesses perceptual similarity by comparing luminance, contrast, and structure between \(\hat{\X}\) and \(\X\) on a band-wise basis. Values close to 1 indicate high structural similarity.
    \[
    \mathrm{SSIM}(\X, \hat{\X}) = \frac{1}{C}\sum_{k=1}^C\frac{(2\mu_k\mu_{\hat{k}} + c_1)(2\sigma_{k\hat{k}} + c_2)}{(\mu_k^2 + \mu_{\hat{k}}^2 + c_1)(\sigma_k^2 + \sigma_{\hat{k}}^2 + c_2)},
    \]
    where \(\mu_k,~\mu_{\hat{k}},~\sigma_k^2,~\sigma_{\hat{k}}^2\), and \(\sigma_{k\hat{k}}\) are $k$-band means, variances, and covariances of $\X$ and $\hat{\X}$, respectively, and \(c_1, c_2\) are small constants to stabilize the denominator.

    \item {\bf Universal Image Quality Index (UIQI)}~\cite{UIQI}: Measures similarity in terms of luminance, contrast, and structure on a band-wise basis. Values range from \(-1\) to 1, with higher values indicating better quality.
    \[
    \mathrm{UIQI}(\X,\hat{\X}) = \frac{1}{C} \sum_{k=1}^C \frac{4\sigma_{k\hat{k}}\mu_k\mu_{\hat{k}}}{\left(\sigma_k^2 + \sigma_{\hat{k}}^2\right)\left(\mu_k^2 + \mu_{\hat{k}}^2\right)}.
    \]

    \item {\bf Erreur Relative Globale Adimensionnelle de Synth\`{e}se (ERGAS)}~\cite{ERGAS}: Provides a global indication of the relative error, normalized by the mean reflectance, and is commonly used in remote sensing. Lower ERGAS indicates higher reconstruction accuracy.
    \[
    \mathrm{ERGAS}(\X,\hat{\X}) = 100~\frac{r}{s}\sqrt{\frac{1}{C}\sum_{k=1}^{C}\frac{\mathrm{RMSE}_k(\X, \hat{\X})^2}{\mu_k^2}},
    \]
    where \(r/s\) is the ratio of spatial resolutions between \(\hat{\X}\) and \(\X\), and \(\mathrm{RMSE}_k(\X, \hat{\X})\) is the RMSE of band \(k\) between $\X$ and $\hat{\X}$.

    \item {\bf Spectral Angle Mapper (SAM)}~\cite{SAM}: Computes the mean spectral angle (in degrees) between estimated and ground-truth spectral vectors at each pixel. It measures spectral similarity, with smaller angles indicating better fidelity. To ensure numerical stability, the arccos argument is clipped to avoid numerical issues.

    \[
    \mathrm{SAM}(\X, \hat{\X}) = \frac{1}{HW} \sum_{i=1}^{H}\sum_{j=1}^{W} \left( \frac{180}{\pi} \arccos\left( \min\left( \frac{\langle \X_{ij:}, \hat{\X}_{ij:} \rangle}{\|\X_{ij:}\|_2 \|\hat{\X}_{ij:}\|_2 + \epsilon}, 1 - \delta \right) \right) \right),
    \]
    
    where $\X_{ij:}$ and $\hat{\X}_{ij:}$ are the spectral vectors at pixel $(i,j)$, and $\epsilon$ and $\delta$ are small constants to avoid division by zero and numerical overflow, respectively (e.g., $\epsilon=10^{-8}$, $\delta=10^{-9}$).

\end{itemize}

\section{Cross-Method Robustness Study}
\label{sec:study}

\subsection{Setup}

We use HyperBench to evaluate six recent HSR methods---MIAE~\cite{MIAE}, C2FF~\cite{C2FF}, SDP~\cite{SDP}, SSSR~\cite{SSSR}, SpectraLift~\cite{SpectraLift}, and SpectraMorph~\cite{SpectraMorph}---on four widely used hyperspectral scenes: Washington DC Mall, Pavia University, Pavia Centre, and Botswana. For each (method, scene) pair, HyperBench produces $70$ reconstructions, formed as the product of the ten PSFs in Section~\ref{subsec:psf} and seven $(r, c)$ operating points, where $r$ denotes the spatial downsampling factor and $c$ the number of MSI bands. The seven operating points, ordered from least to most demanding along the spatial axis, are:
\begin{itemize}
    \setlength\itemsep{1pt}
    \item $(r=4,\;c=4)$ at LR-HSI SNR $= 35$\,dB --- base configuration;
    \item $(r=8,\;c=3)$, $(r=8,\;c=4)$, $(r=8,\;c=8)$, and $(r=8,\;c=16)$ at SNR $= 30$\,dB;
    \item $(r=16,\;c=4)$ at SNR $= 25$\,dB;
    \item $(r=32,\;c=4)$ at SNR $= 20$\,dB --- most demanding configuration.
\end{itemize}
These seven points jointly sweep the spatial-resolution axis ($r \in \{4, 8, 16, 32\}$ at fixed $c=4$) and the spectral-richness axis ($c \in \{3, 4, 8, 16\}$ at fixed $r=8$), sharing the $(r=8, c=4)$ anchor. The LR-HSI SNR is matched to the downsampling factor in $5$\,dB steps to reflect the photon-budget reduction inherent to coarser-resolution observations; HR-MSI noise is held fixed at SNR $= 40$\,dB throughout. The SRFs at $c \in \{3, 4\}$ are derived from IKONOS, $c=8$ from WorldView-2, and $c=16$ from WorldView-3, as described in Section~\ref{subsec:srf}.

Across the six methods, the grid yields $6 \times 4 \times 70 = 1{,}680$ reconstructions. Each value reported in Figures~\ref{fig:robustness-psnr} and~\ref{fig:robustness-sam} and in Tables~\ref{tab:psf-psnr} and~\ref{tab:psf-sam} is an average over every factor that does not appear on the corresponding axis; error bars and standard deviations are taken over those same factors. The objective is to characterise the dependence of each method's reported quality on the choice of degradation operators---a dependence that is structurally hidden by the prevailing single-configuration evaluation protocol.

\begin{table}[!t]
\centering
\footnotesize
\caption{Mean PSNR (dB; higher is better) per (method, PSF), averaged over the four scenes and the seven HyperBench configurations per PSF (28 measurements per cell). Best PSNR per row in \textbf{bold}.}
\label{tab:psf-psnr}
\begin{tabular}{|l|c|c|c|c|c|c|}
\hline
PSF & MIAE & C2FF & SDP & SSSR & SpectraLift & SpectraMorph \\
\hline
Hermite & 35.51 & 35.23 & 34.06 & 32.63 & {\bf 36.47} & 36.35 \\
\hline
Gabor & 35.27 & 35.35 & 34.12 & 32.42 & {\bf 36.55} & 36.37 \\
\hline
Gaussian & 34.84 & 35.59 & 34.16 & 31.99 & {\bf 36.57} & 36.45 \\
\hline
Lorentz$^2$ & 34.05 & 35.81 & 34.18 & 31.19 & {\bf 36.62} & 36.49 \\
\hline
Kolmog. & 33.72 & 35.78 & 34.25 & 31.23 & {\bf 36.66} & 36.51 \\
\hline
Moffat & 31.39 & 35.80 & 34.22 & 29.20 & {\bf 36.85} & 36.67 \\
\hline
Parabolic & 30.73 & 36.16 & 34.22 & 28.49 & {\bf 36.91} & 36.68 \\
\hline
Airy & 28.50 & 36.09 & 33.96 & 26.39 & {\bf 36.92} & 36.77 \\
\hline
Delta & 27.01 & 36.01 & 33.81 & 24.09 & {\bf 37.03} & 36.88 \\
\hline
Sinc & 26.46 & 36.27 & 33.13 & 23.96 & {\bf 36.99} & 36.87 \\
\hline
\end{tabular}
\end{table}

\begin{table}[!t]
\centering
\footnotesize
\caption{Mean SAM (degrees; lower is better) per (method, PSF), averaged over the four scenes and the seven HyperBench configurations per PSF (28 measurements per cell). Best SAM per row in \textbf{bold}.}
\label{tab:psf-sam}
\begin{tabular}{|l|c|c|c|c|c|c|}
\hline
PSF & MIAE & C2FF & SDP & SSSR & SpectraLift & SpectraMorph \\
\hline
Hermite & 3.06 & 3.13 & 3.80 & 3.51 & 2.85 & {\bf 2.78} \\
\hline
Gabor & 3.17 & 3.18 & 3.77 & 3.58 & 2.80 & {\bf 2.79} \\
\hline
Gaussian & 3.30 & 3.09 & 3.75 & 3.67 & 2.81 & {\bf 2.76} \\
\hline
Lorentz$^2$ & 3.65 & 2.99 & 3.71 & 3.83 & 2.79 & {\bf 2.73} \\
\hline
Kolmog. & 3.82 & 2.99 & 3.69 & 3.81 & 2.78 & {\bf 2.73} \\
\hline
Moffat & 4.87 & 3.04 & 3.66 & 4.28 & 2.70 & {\bf 2.69} \\
\hline
Parabolic & 5.19 & 2.93 & 3.66 & 4.56 & {\bf 2.66} & 2.68 \\
\hline
Airy & 6.42 & 2.94 & 3.69 & 5.26 & 2.69 & {\bf 2.66} \\
\hline
Delta & 7.23 & 2.94 & 3.67 & 6.16 & {\bf 2.64} & 2.65 \\
\hline
Sinc & 7.71 & 2.94 & 3.81 & 6.04 & {\bf 2.64} & {\bf 2.64} \\
\hline
\end{tabular}
\end{table}

\subsection{Cross-PSF Behavior}

\begin{figure}[!htbp]
    \centering
    \includegraphics[width=\textwidth]{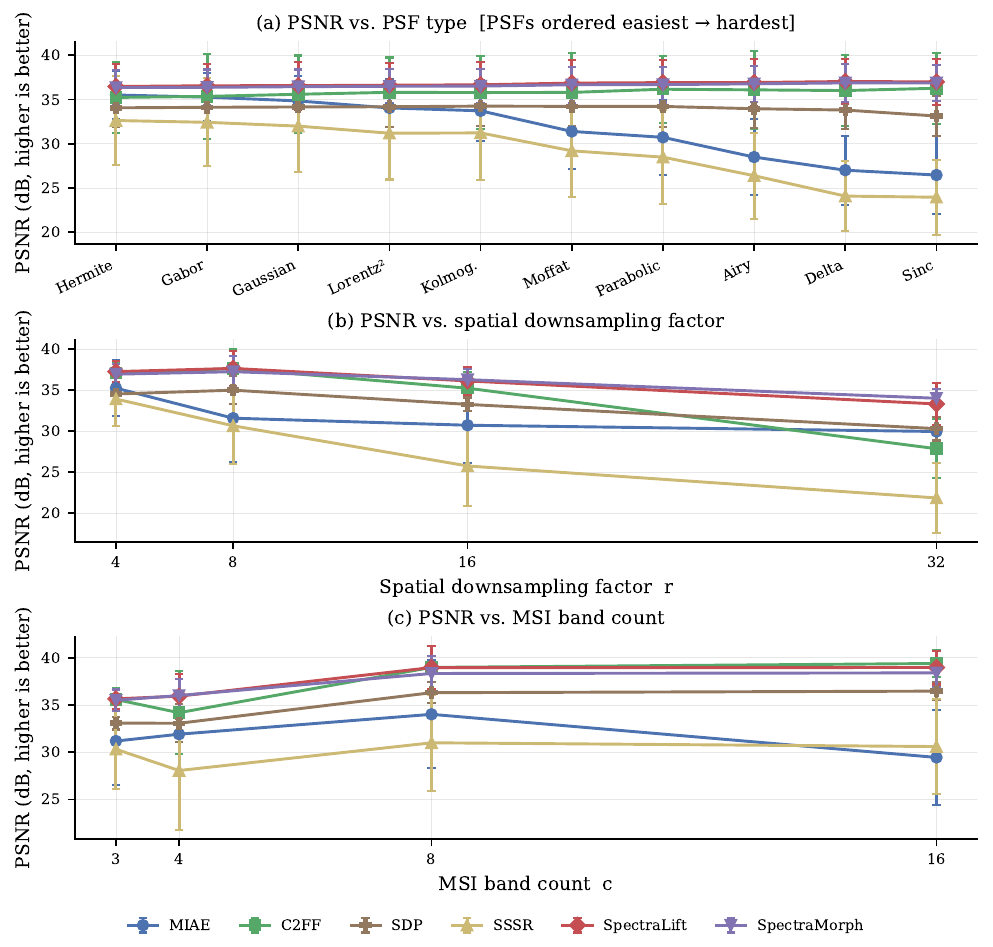}
    \caption{PSNR (dB; higher is better) of six HSR methods evaluated with HyperBench. \textbf{(a)}~PSNR per PSF, averaged over the four scenes and the seven $(r,c)$ operating points (28 measurements per cell). PSFs are ordered along the $x$-axis from easiest to hardest based on mean PSNR across all six methods. \textbf{(b)}~PSNR per spatial downsampling factor, averaged over the four scenes, the ten PSFs, and every $c$ value present at that $r$ in the experiment grid (the $r=8$ point pools $c \in \{3, 4, 8, 16\}$; the other $r$ levels are evaluated only at $c=4$). \textbf{(c)}~PSNR per MSI band count, averaged over the four scenes, the ten PSFs, and every $r$ value present at that $c$ (the $c=4$ point pools $r \in \{4, 8, 16, 32\}$; the other $c$ levels are evaluated only at $r=8$). Error bars span one standard deviation across the aggregated measurements.}
    \label{fig:robustness-psnr}
\end{figure}

\begin{figure}[!htbp]
    \centering
    \includegraphics[width=\textwidth]{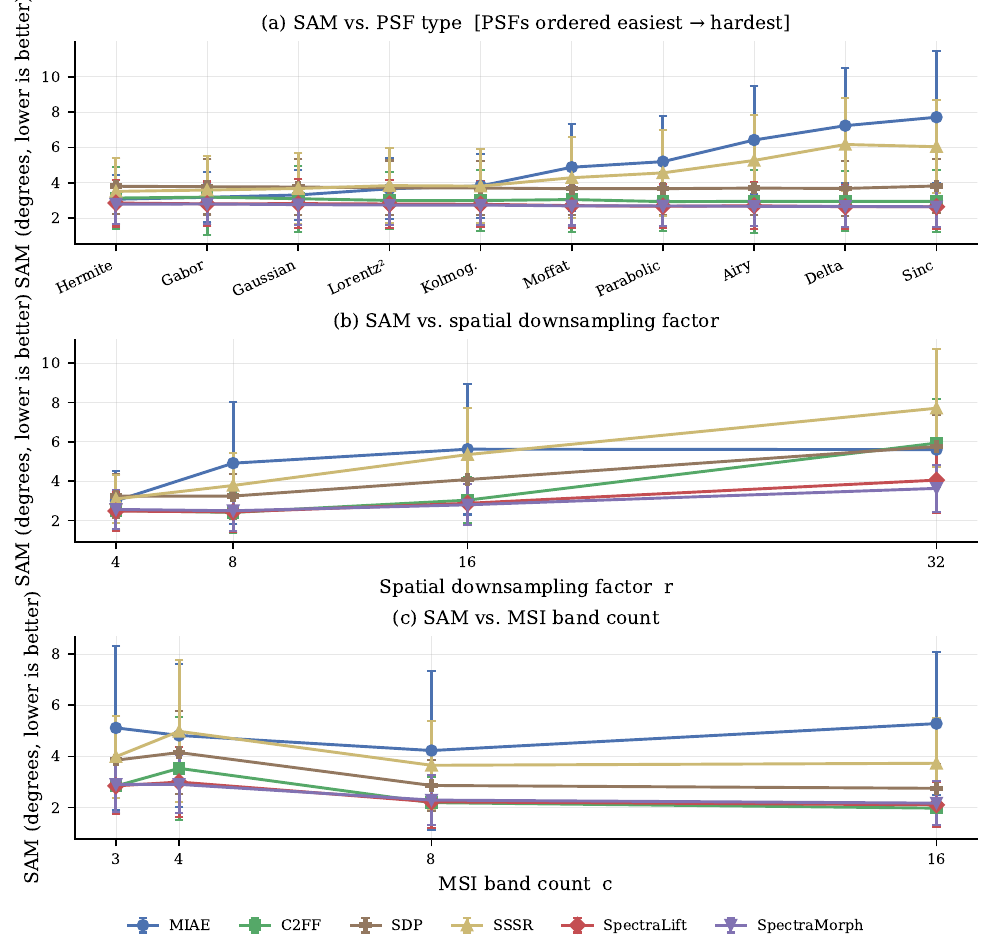}
    \caption{SAM (degrees; lower is better) of six HSR methods evaluated with HyperBench. Layout and aggregation match Figure~\ref{fig:robustness-psnr}. \textbf{(a)}~SAM per PSF, averaged across 28 measurements per cell. \textbf{(b)}~SAM per spatial downsampling factor, averaged as in Figure~\ref{fig:robustness-psnr}(b). \textbf{(c)}~SAM per MSI band count, averaged as in Figure~\ref{fig:robustness-psnr}(c).}
    \label{fig:robustness-sam}
\end{figure}

The line plot in panel~(a) of Figure~\ref{fig:robustness-psnr} reveals a clear bifurcation in PSF robustness. Across the five best-performing PSFs (Hermite, Gabor, Gaussian, Lorentz$^2$, Kolmogorov), the six methods lie within an approximately $5$\,dB band ($31.2$--$36.7$\,dB): the same band within which state-of-the-art claims are routinely made and contested in the literature. Across the five remaining PSFs (Moffat, Parabolic, Airy, Delta, Sinc), the methods split into two clusters: for some methods MIAE drops by approximately $9$\,dB between Hermite ($35.5$\,dB) and Sinc ($26.5$\,dB), and SSSR drops by a comparable margin ($32.6 \to 24.0$\,dB), while for others each remain within $1.5$\,dB of their best-PSF score on every PSF family. The inter-method PSNR spread thus widens from approximately $4$\,dB on the easiest PSF to over $13$\,dB on the hardest -- a more than threefold expansion attributable entirely to the choice of degradation operator. Panel~(b) exposes an analogous spatial-axis fragility: between $r=4$ and $r=32$, SSSR loses approximately $12$\,dB and C2FF approximately $9$\,dB, while SDP, SpectraLift, and SpectraMorph lose only $3$--$4$\,dB. Panel~(c) shows a structural under-performance of MIAE across all MSI band counts, consistent with the weaker spectral fidelity quantified in Figure~\ref{fig:robustness-sam}.

Figure~\ref{fig:robustness-sam} provides a complementary spectral picture. SAM measures the per-pixel angular distance between predicted and ground-truth spectra and is therefore sensitive to spectral distortions that PSNR may mask. Across the easiest five PSFs, the six methods lie within an approximately $1.0^\circ$ band ($2.7^\circ$--$3.8^\circ$). Across the hardest five PSFs, MIAE's SAM grows by a factor of approximately $2.5$ (from $3.1^\circ$ on Hermite to $7.7^\circ$ on Sinc) and SSSR's by a factor of approximately $1.7$ (from $3.5^\circ$ to $6.0^\circ$), while C2FF, SDP, SpectraLift, and SpectraMorph each remain within $0.4^\circ$ of their best-PSF spectral fidelity throughout. The inter-method SAM spread widens from approximately $0.7^\circ$ on the easiest PSF to over $5^\circ$ on the hardest---a sevenfold expansion. The same bifurcation is visible along the spatial-downsampling axis (panel~b), where at $r=32$ SSSR's SAM is more than twice that of SpectraMorph.

The per-(method, PSF) values that underlie panel~(a) of both figures are reported in Tables~\ref{tab:psf-psnr} and~\ref{tab:psf-sam}; cells in bold indicate the best (highest PSNR, lowest SAM) value per row.

\subsection{Per-Method Observations}

\noindent\textbf{MIAE}~\cite{MIAE}\textbf{.}~Competitive on PSFs whose spatial profile is smooth and well-localised (Hermite, Gabor, Gaussian, Lorentz$^2$, Kolmogorov: PSNR $\geq 33.7$\,dB; SAM $\leq 3.8^\circ$), but degrades sharply on PSFs whose impulse response is either oscillatory (Sinc, Airy) or absent (Delta, where downsampling aliasing dominates), where PSNR drops to $26.5$--$28.5$\,dB and SAM rises to $6.4$--$7.7^\circ$. The pattern is consistent with an inversion that implicitly assumes a smooth, spatially localised PSF and therefore fails to extrapolate to kernels that violate that assumption.

\noindent\textbf{C2FF}~\cite{C2FF}\textbf{.}~Strongly PSF-invariant: PSNR remains within $35.2$--$36.3$\,dB and SAM within $2.9$--$3.2^\circ$ across all ten PSF families. The cost of this PSF invariance appears along the spatial-resolution axis, where PSNR decreases by approximately $9$\,dB between $r=4$ and $r=32$---the steepest spatial collapse among the four PSF-robust methods (C2FF, SDP, SpectraLift, SpectraMorph). The pattern suggests an inversion that is robust to kernel shape but sensitive to information loss at high downsampling.

\noindent\textbf{SDP}~\cite{SDP}\textbf{.}~The most PSF-invariant of the six methods in spectral fidelity: SAM remains within a $0.15^\circ$ window ($3.66$--$3.81^\circ$) across all ten PSF families---a tighter cross-PSF range than any other measured method. PSNR is similarly stable, varying by only $\sim 1.1$\,dB ($33.1$--$34.3$\,dB) across PSFs and dropping by only $\sim 4$\,dB between $r=4$ and $r=32$. SDP's absolute spatial and spectral fidelity sit consistently below SpectraLift and SpectraMorph by $1$--$3$\,dB PSNR and $\sim 1^\circ$ SAM, but it never approaches the collapse exhibited by MIAE and SSSR on the hardest PSFs.

\noindent\textbf{SSSR}~\cite{SSSR}\textbf{.}~The most fragile of the six methods along every axis examined. PSNR decreases by approximately $9$\,dB between the easiest and hardest PSFs and by approximately $12$\,dB between $r=4$ and $r=32$; SAM increases by approximately $70$\,\% across the hardest PSFs and by more than $150$\,\% across the largest downsampling factor. Under the single-Gaussian, $r=4$ protocol, SSSR sits within $4$--$5$\,dB of the strongest methods; under HyperBench's broader sweep, it is consistently the lowest-performing method.

\noindent\textbf{SpectraLift}~\cite{SpectraLift}\textbf{.}~Highest PSNR on every PSF family (narrowly) and best or tied-best SAM on the three most challenging PSFs (Parabolic, Sinc, Delta). Retains approximately $33$\,dB PSNR at $r=32$.

\noindent\textbf{SpectraMorph}~\cite{SpectraMorph}\textbf{.}~Best SAM on seven of the ten PSFs and PSNR within $0.1$\,dB of SpectraLift on every PSF. The most spatially robust method evaluated: PSNR decreases by only approximately $3$\,dB between $r=4$ and $r=32$.

\section{Discussion}

\subsection{Implications of the Cross-Method Findings}

The results in Section~\ref{sec:study} have direct consequences for how the HSR field should report and interpret reconstruction quality. Most concretely, the inter-method PSNR spread varies by more than threefold across the PSF axis alone: methods that lie within $5$\,dB of one another on the easiest PSF separate by more than $13$\,dB on the hardest. Reported state-of-the-art figures are therefore best understood as configuration-conditional rather than absolute, and direct numerical comparisons across papers that fix different configurations are misleading even when the methods are nominally evaluated under the same protocol. None of the individual papers examined here has made false claims; rather, each paper's results are valid for the small subset of the degradation space that paper happened to evaluate, and the convention of reporting at a single point in that space is a field-level convention rather than an author-level choice.

The bifurcation between PSF-fragile methods (MIAE, SSSR) and PSF-robust methods (C2FF, SDP, SpectraLift, SpectraMorph) is plausibly diagnostic of modeling-assumption mismatch rather than an unrecoverable property of any individual architecture. Within the PSF-robust cluster, the methods further differentiate along the spatial-resolution axis: SpectraLift, SpectraMorph, and SDP retain their absolute fidelity even at $r=32$, while C2FF degrades by approximately $9$\,dB---a pattern consistent with an inversion that is robust to kernel shape but sensitive to information loss at high downsampling. Exposing such assumptions through systematic sweeps is itself a contribution that the conventional single-configuration protocol cannot make, and it points toward a more constructive use of broad evaluation: not as an instrument for ranking but as a diagnostic for the regimes under which each modeling choice is defensible.

The appropriate response to these findings is not to crown a winner based on a single aggregate score, but to make robustness reporting a routine component of HSR evaluation. A method's reconstruction-quality summary should include the configuration grid over which it was evaluated, the spread within that grid, and the regimes under which it is reliable. HyperBench is the operational mechanism through which this becomes a low-friction practice rather than a time-consuming undertaking.

\subsection{HyperBench as Community Infrastructure}

Beyond enabling individual robustness studies, HyperBench is intended to function as shared community infrastructure for HSR evaluation. The framework's value as such rests on three properties that are difficult to provide in any single-author research artifact: comparability, accessibility, and extensibility.

Comparability follows from the framework's pipeline being a standardized and versioned artifact rather than a per-paper reimplementation. When two methods are evaluated within HyperBench, they share identical data generation, normalization, SRF construction, noise injection, spatial downsampling, and metric computation. Performance differences therefore reflect the methods themselves rather than incidental implementation choices---normalization conventions, boundary handling, noise seeds, or SRF construction methodology---that have historically introduced silent variance into cross-paper comparisons.

Accessibility follows from the decoupling of model code from experimental pipeline. A researcher developing a new reconstruction method need not also design and validate a complete evaluation pipeline; they implement against a single interface and the framework handles the rest. This separation of concerns is particularly valuable for researchers entering HSR from adjacent fields---satellite imaging, medical hyperspectral analysis, agricultural remote sensing---who wish to evaluate methods on their domain data without investing in bespoke evaluation infrastructure.

Extensibility follows from HyperBench's modular design. New degradation models, sensor configurations, evaluation metrics, and dataset loaders can be added without disrupting existing experiments. As new sensing modalities become relevant, the degradation library can grow correspondingly, providing a stable evaluation substrate that remains compatible across an evolving research landscape. \textbf{HyperBench is designed as an extensible benchmarking framework rather than a closed benchmark. We encourage community contributions of sensor-specific PSFs, SRFs, degradation operators, evaluation metrics, and baseline implementations, which can broaden the framework’s coverage and improve the realism and reproducibility of HSI-MSI fusion evaluation.}

\section{Conclusion}

We have presented HyperBench, a framework for standardized, scalable, and reproducible synthetic evaluation of hyperspectral super-resolution methods. HyperBench specifies Wald's protocol with a comprehensive and configurable degradation pipeline spanning ten point spread functions, four spectral response function configurations derived from operational multispectral sensors, arbitrary spatial downsampling factors, and adjustable noise levels. By decoupling experimental design from model development, the framework enables fair cross-method comparison, automated large-scale benchmarking, and precise targeted experimentation within a single infrastructure.

A cross-method study conducted with HyperBench (Section~\ref{sec:study}) provides empirical grounding for the framework's motivation. The inter-method PSNR spread among six recent HSR methods widens by more than threefold across the PSF axis alone, and the methods bifurcate into a PSF-fragile cluster (MIAE, SSSR) that loses $8$--$12$\,dB and sees SAM grow by a factor of $2$--$3$ under non-Gaussian PSFs, and a PSF-robust cluster (C2FF, SDP, SpectraLift, SpectraMorph) whose reconstruction quality remains within $1.5$\,dB and $0.4^\circ$ of its best-PSF value throughout. Findings of this character are, by construction, invisible to the single-configuration evaluation protocols that dominate the published HSR literature; they become routine under HyperBench because the entire sweep is specified by a single configuration file.

The release of HyperBench as an open-source framework is intended to serve as a shared evaluation substrate for the HSR community. The framework is designed to grow with the field, incorporating new degradation models, datasets, and metrics as the community's understanding of realistic sensing conditions deepens. The objective is not only to improve the rigor of individual evaluations but to establish the conditions under which the field's collective progress can be measured and trusted.

\section*{Acknowledgement}
We thank the authors of \cite{MIAE,C2FF,SDP,SSSR} for sharing implementations of their proposed methods with the research community, and we thank Mario Parente for many helpful comments and suggestions.

\clearpage

\printbibliography

\end{document}